\documentstyle[12pt,epsf,]{article}

\newcommand{\beq}{\begin{equation}}
\newcommand{\eeq}{\end{equation}}
\newcommand{\beqa}{\begin{eqnarray}}
\newcommand{\eeqa}{\end{eqnarray}}
\newcommand{\no}{\nonumber}
\newcommand{\q}{\quad}
\newcommand{\qq}{\qquad}
\newcommand{\mnod}{\stackrel{\circ}{M}}
\newcommand{\Fnod}{\stackrel{\circ}{F}}
\newcommand{\tr}{\mbox{tr}}

\begin{document}

\hfill 

\hfill 

\bigskip\bigskip

\begin{center}

{{\Large\bf Sigma--terms in heavy baryon chiral perturbation theory revisited
     }}

\end{center}

\vspace{.4in}

\begin{center}
{\large B. Borasoy\footnote{email: borasoy@het.phast.umass.edu }}

\bigskip

\bigskip

Department of Physics and Astronomy\\
University of Massachusetts\\
Amherst, MA 01003, USA\\

\vspace{.2in}

\end{center}

\vspace{.7in}

\thispagestyle{empty} 

\begin{abstract}
The $\sigma$--terms are calculated at next--to--leading order
in heavy baryon chiral perturbation
theory by employing a cutoff regularization. The results do not depend on 
the cutoff value to the order we are working . The baryon masses
and $\sigma_{\pi N}(0)$ are used to perform a least--squares fit
to the three appearing low--energy constants and predictions for the
two $KN$ $\sigma$--terms and the strange contribution to the nucleon mass
are made. The lack of convergence in the chiral expansions of
these quantities when regularized dimensionally is overcome in the
cutoff scheme.
The $\sigma$--term shifts to the pertinent Cheng--Dashen points are
calculated.
We also include the spin--3/2 decuplet in the effective theory.
\end{abstract}

\vspace{.7in}


\vfill

\section{Introduction}
Chiral perturbation theory  which is the effective field theory of the 
Standard Model at low energies in the hadronic sector has been successfully 
applied within the sector of Goldstone Bosons \cite{GL}. However,
traditional SU(3) heavy baryon chiral perturbation theory does not appear 
to work well. The leading nonanalytic components from loop corrections
destroy the good experimental agreement which exists at lowest order.
The additional contributions have to be compensated by higher 
order countertems. This leads to problems with the convergence of the 
chiral series.

Recently a resolution of this problem was proposed by using a cutoff
regularization instead of the common dimensional regularization scheme
\cite{DH,DHB}. There, the authors come to the conclusion that
dimensionally regularized Feynman diagrams carry implicit and large
contributions from short distance physics. In contrast, the
cutoff scheme picks out the long distance part of the integral, which
behaves as expected on physical grounds. Both an exponential cutoff
in three--momentum and a dipole regulator were employed therein.
However, in the cases discussed there the cutoff is irrelevant 
-- a consistent chiral expansion can then be carried 
out.

In these works an analysis of the octet baryon masses has been given
using the lowest and next--to--leading order in the derivative expansion
of the effective Lagrangian. Further information on the scalar sector
of baryon CHPT is given by the scalar form factors or $\sigma$--terms 
which measure the strength of the various matrix elements
$m_q \, \bar{q} q \: (q = u,d,s)$ in the proton and vanish in the
chiral limit of zero quark masses. Thus, they are particularly suited
to test our understanding of spontaneous and explicit chiral
symmetry breaking.
The purpose of this work is to examine these quantities employing
cutoff regularization schemes.

Another complication arises from the closeness of the spin--3/2 decuplet
resonances which are separated only by 231 MeV in average from the octet
baryons which is considerably smaller than the kaon and the $\eta$ mass.
These resonances are, therefore, expected to play an important part
at low energies. It has been suggested \cite{JM} to include the
decuplet explicitely.

The present work is organized as follows. 
In the next section we apply two different cutoff schemes to regularize
the Feynman diagrams without decuplet contributions.
Besides the dipole cutoff already used in
\cite{DH,DHB} we consider a slightly modified dipole cutoff which is 
identical to the first one for vanishing off--shell momenta of the baryons.
The following section
deals with the inclusion of the decuplet fields. 
The results for both cases are presented in Sec.~4. In Sec.~5 we conclude
with a short 
summary. The decuplet contributions to the scalar form factors and the
octet baryon masses are relegated
to the Appendices.

\section{$\sigma$--terms in a cutoff scheme}
In this section, we will work with the heavy baryon Lagrangian
for the Goldstone bosons and the octet baryons
which can be decomposed into a lowest order and a next--to--leading order
part in the derivative expansion
\beq
{\cal L}_{\phi \, B} = {\cal L}_{\phi \, B}^{(1)} 
                     + {\cal L}_{\phi \, B}^{(2)}  \qq ,
\eeq
where the superscript denotes the chiral order.
In the heavy baryon formulation the baryons are described by a 
four--velocity $v_{\mu}$ and relativistic corrections appear as
$1/ \mnod$ corrections where $\mnod$ is the average octet baryon mass
in the chiral limit. A consistent chiral counting scheme emerges,
${\it i.e.}$ a one--to--one correspondence between the Goldstone boson loops
and the expansion in small momenta and quark masses.
The lowest order Lagrangian $ {\cal L}_{\phi \, B}^{(1)}$ includes the 
two axial--vector couplings $D$ and $F$
\beq
{\cal L}_{\phi \, B}^{(1)} = \, i \, \tr \Big( \bar{B} 
       [ v \cdot D , B] \Big) + 
   D \, \tr \Big( \bar{B} S_{\mu} \{ u^{\mu}, B\} \Big) 
  + F \, \tr \Big( \bar{B} S_{\mu} [ u^{\mu}, B] \Big) 
\eeq
where $ 2 S_{\mu} = i \gamma_5 \sigma_{\mu \nu} v^{\nu} $ denotes the
Pauli--Lubanski spin vector.
The pseudoscalar Goldstone fields ($\phi = \pi, K, \eta$) are collected in
the  $3 \times 3$ unimodular, unitary matrix $U(x)$, 
\begin{equation}
 U(\phi) = u^2 (\phi) = \exp \lbrace 2 i \phi / \Fnod \rbrace
\end{equation}
with $\Fnod$ being the pseudoscalar decay constant (in the chiral limit), and
\begin{eqnarray}
 \phi =  \frac{1}{\sqrt{2}}  \left(
\matrix { {1\over \sqrt 2} \pi^0 + {1 \over \sqrt 6} \eta
&\pi^+ &K^+ \nonumber \\
\pi^-
        & -{1\over \sqrt 2} \pi^0 + {1 \over \sqrt 6} \eta & K^0
        \nonumber \\
K^-
        &  \bar{K^0}&- {2 \over \sqrt 6} \eta  \nonumber \\} 
\!\!\!\!\!\!\!\!\!\!\!\!\!\!\! \right) \, \, \, \, \, . 
\end{eqnarray}
Under SU(3)$_L \times$SU(3)$_R$, $U(x)$ transforms as $U \to U' =
LUR^\dagger$, with $L,R \in$ SU(3)$_{L,R}$.
One forms an object of axial--vector type with one derivative
\beq
u_{\mu} = i u^{\dagger} \nabla_{\mu} U u^{\dagger}
\eeq
with $\nabla_{\mu}$ being the covariant derivative of $U$. 
The matrix $B$ denotes the baryon octet, 
\begin{eqnarray}
B  =  \left(
\matrix  { {1\over \sqrt 2} \Sigma^0 + {1 \over \sqrt 6} \Lambda
&\Sigma^+ &  p \nonumber \\
\Sigma^-
    & -{1\over \sqrt 2} \Sigma^0 + {1 \over \sqrt 6} \Lambda & n
    \nonumber \\
\Xi^-
        &       \Xi^0 &- {2 \over \sqrt 6} \Lambda \nonumber \\} 
\!\!\!\!\!\!\!\!\!\!\!\!\!\!\!\!\! \right)  \, \, \, .
\end{eqnarray}
The matrices $u_{\mu}$ and $B$ transform under $SU(3)_L \times SU(3)_R$ 
as any matter field, ${\it e.g.}$,
\begin{equation} 
B \to B' = K \, B \,  K^\dagger
 \, \, \, ,
\end{equation}
with $K(U,L,R)$ the compensator field representing an element of the
conserved subgroup SU(3)$_V$.
We will use $D = 3/4$ and $F = 1/4$ which leads to $g_A = D + F = 1.25$.
At next--to--leading order explicit chiral symmetry breaking terms appear
\beq
{\cal L}_{\phi \, B}^{(2)} = \, b_0 \, \tr \Big( \bar{B} B \Big)  \,
            \tr \Big( \chi_+ \Big) \,  + \,
         b_D \, \tr \Big( \bar{B} \{ \chi_+ , B \} \Big)  \, + \,
         b_F \, \tr \Big( \bar{B} [ \chi_+ , B ] \Big)   
\eeq
with $\chi_+ =  2 B_0 \, ( u^{\dagger} {\cal M} u^{\dagger} + u {\cal M} u $)
and ${\cal M} = \mbox{diag} ( m_u,m_d,m_s )$ the quark mass matrix.
We prefer to work in the isospin limit $m_u = m_d = \hat{m}$.
To this order three coupling constant, so called low--energy 
constants (LECs), appear. 
Together with the average octet baryon mass in the chiral limit $\mnod$ 
we end up with four unknown parameters.
These have to be fixed from phenomenology.
Here we will use the four different masses for the octet baryons in the
isospin limit. Since both $\mnod$ and $b_0$ shift the baryon mass
spectrum by a constant the pion nucleon sigma--term at zero momentum 
transfer $\sigma_{\pi N} (0)$ ( or one of the kaon nucleon sigma--terms )
has also to be taken into account in order to fix all parameters.

One defines the scalar form factors or $\sigma$--terms which measure
the strength of $m_q \bar{q} q$ in the proton by
\beqa
\sigma_{\pi N} (t) & = & \hat m \, <p' \, | \bar u u + \bar d d| \, p>
\, \, \, , \nonumber \\
\sigma_{KN}^{(1)} (t) & = & \frac{1}{2}(\hat m + m_s) \, 
<p' \, | \bar u u + \bar s s| \, p> \, \, \, , \nonumber \\
\sigma_{KN}^{(2)} (t) & = & \frac{1}{2}(\hat m + m_s) \, 
<p' \, | -\bar u u + 2\bar d d + \bar s s| \, p> \, \, \, , \no 
\eeqa
with $| \, p>$ a proton state with four--momentum $p$, 
$t = (p'-p)^2$ the invariant momentum transfer squared.
It is most convenient to work in the Breit--frame in which
$v \cdot p' = v \cdot p$.
There are two types of contributions for the $\sigma$--terms.
To lowest order, there are the tree level contributions of chiral order two
from the counterterms $b_0, b_D, b_F$ of the Lagrangian 
${\cal L}_{\phi B}^{(2)}$. The contributions from the Goldstone boson loops
appear at next--to--leading order and
can be evaluated by using dimensional regularization \cite{BKM}.
A typical integral in this analysis has for the case
of zero momentum transfer and $d$ dimensions the form
\beq  \label{int}
\int  \frac{ d^d l }{(2 \pi)^d} \, \, \frac{ i^3 \, ( S \cdot l )^2}{
       [l^2 - m_{\phi}^2 + i \epsilon]^2 \, [ v \cdot l + i \epsilon ]} \:
    =  \: \frac{3}{64 \pi} \, m_{\phi} 
\eeq
where $m_{\phi}$ is the meson mass.
The result is
non--analytic in the quark masses since $m_{\phi} \propto m_q^{1/2}$.
The integral grows linearly with increasing meson mass.
We expect the long distance portion of the integral to be larger
for small meson masses since for small momenta the meson propagator
can be approximated by $1/m_{\phi}^2$.
This indicates that in the dimensionally regularized integral
there are significant contributions from short distance physics
which cannot be described appropriately by chiral symmetry.
Therefore, one has to employ other regularization schemes that emphasize
long distance effects of the integrals and reduce short distance
contributions. In \cite{DHB} it was shown that a simple dipole
regulator fulfills these requirements.

For the evaluation of the Goldstone boson loops we will employ the regulators
\beq
R_1 = \bigg( \frac{ \Lambda^2}{\Lambda^2 - l^2} \bigg)^2  \q , \q
R_2 = \bigg( \frac{ \Lambda^2}{\Lambda^2 -l^2} \bigg )
      \bigg( \frac{ \Lambda^2}{\Lambda^2 - (l+q)^2} \bigg) 
\eeq
where $l$ is the loop momentum and $q$ the small off--shell momentum
of the external baryons in the heavy mass formalism.
Both cases are identical for vanishing off--shell momentum $q$. 
The reason for considering the regulator $R_2$ in addition to $R_1$
will become clear when we introduce decuplet fields. It turns out that
including decuplet fields leads to divergent integrals  in the case of the
regulator $R_1$, whereas $R_2$ avoids this difficulty.
Also, we are able to compare the results for the $\sigma$--terms in both
regularization schemes and examine the dependence on employing
different regulators.
Inserting these regulators into the integral in Eq.~(\ref{int}) leads to
\beq  \label{intlam}
I_{\Lambda} =
\int  \frac{ d^d l }{(2 \pi)^d} \, \, \frac{ i^3 \, ( S \cdot l )^2}{
       [l^2 - m_{\phi}^2 + i \epsilon]^2 \, [ v \cdot l + i \epsilon ]}
       \bigg( \frac{ \Lambda^2}{\Lambda^2 - l^2} \bigg)^2  \:
  = \:  - \frac{1}{64 \pi} \frac{ \Lambda^4}{ (\Lambda + m_{\phi})^3 } \qq .
\eeq
The introduction of the additional scale $\Lambda$ spoilt the
one--to--one correspondence between the meson loops and the expansion
in the quark masses
and the integral depends strongly on the value of the cutoff $\Lambda$.
However, this does not mean that to the order
we are working the resulting physics will depend
on $\Lambda$, since one is able to absorb the effects of $\Lambda$ into
a renormalization of the LECs.
To this end, one expands the result in Eq.~(\ref{intlam}) in terms 
of the meson mass $m_{\phi}$
\beq
I_{\Lambda}  \stackrel{m_{\phi}<<\Lambda}{\longrightarrow}
- \frac{1}{64 \pi}\Lambda + \frac{3}{64 \pi} m_{\phi} + \ldots
\eeq
where the ellipsis stands for higher orders in $m_{\phi}$.
The second term in the expansion agrees with the result from the dimensionally
regularized version. The first term is a constant and can be absorbed
into renormalizations of the coefficients $b_0, b_D, b_F$ of the Lagrangian 
${\cal L}_{\phi B}^{(2)}$, 
and indeed this is found to be the case -- one
verifies that
\begin{eqnarray}
b_D^r &=& b_D - {3 F^2- D^2 \over 128 \pi \Fnod^2} \, \Lambda \nonumber\\
b_F^r &=& b_F - {5 D F \over 192 \pi \Fnod^2} \, \Lambda \nonumber\\
b_0^r &=& b_0 - {13 D^2 + 9 F^2 \over 576 \pi \Fnod^2} \, \Lambda
\end{eqnarray}
which agrees with the result already obtained in the analysis
for the baryon masses \cite{DHB}.
That this renormalization can occur involves a highly constrained set
of conditions and the fact that they are satisfied is a significant
verification of the chiral invariance of the cutoff procedure.

The chiral expansions at next--to--leading order for the $\sigma$--terms read
for zero momentum transfer
\beqa
\sigma_{\pi N} (0)  & = &  \frac{ m_{\pi}^2 }{ 64 \pi F_{\pi}^2 } \,
         \Lambda^4 \, \bigg( 3 [ D+F ]^2 \frac{1}{ (\Lambda+m_{\pi})^3 }
         + \frac{1}{3} [ 5D^2-6DF+9F^2 ] \frac{1}{ (\Lambda+m_K)^3 }  \no \\
 & &    + \frac{1}{9} [ D-3F ]^2 \frac{1}{ (\Lambda+m_{\eta})^3 } \bigg) \:
       -2 m_{\pi}^2 \Big( b_D + b_F + 2 b_0 \Big) \\
\sigma_{K N}^{(1)} (0)  & = &  \frac{ m_{K}^2 }{ 64 \pi F_{\pi}^2 } \,
    \Lambda^4 \, \bigg( \frac{3}{2} [ D+F ]^2 \frac{1}{ (\Lambda+m_{\pi})^3 }
         +  [ \frac{7}{3}D^2-2DF+5F^2 ] \frac{1}{ (\Lambda+m_K)^3 } \no \\
 & &    + \frac{5}{18} [ D-3F ]^2 \frac{1}{ (\Lambda+m_{\eta})^3 } 
  - \frac{1}{3} [D-3F] \, [D+F] \frac{ (m_{\pi} + m_{\eta}) \Lambda
         + 2 m_{\pi} m_{\eta} }{ [\Lambda + m_{\pi}]^2 \,
          [\Lambda + m_{\eta}]^2 \, [ m_{\pi} + m_{\eta} ] } \bigg) \no \\
 & &     - 4 m_K^2 \Big( b_D +  b_0 \Big) \\
\sigma_{K N}^{(2)} (0)  & = &  \frac{ m_{K}^2 }{ 64 \pi F_{\pi}^2 } \,
   \Lambda^4 \, \bigg( \frac{3}{2} [ D+F ]^2 \frac{1}{ (\Lambda+m_{\pi})^3 }
         + 3 [ D - F ]^2 \frac{1}{ (\Lambda+m_K)^3 } \no \\
 & &    + \frac{5}{18} [ D-3F ]^2 \frac{1}{ (\Lambda+m_{\eta})^3 } 
       +  [D-3F] \, [D+F] \frac{ (m_{\pi} + m_{\eta}) \Lambda
         + 2 m_{\pi} m_{\eta} }{ [\Lambda + m_{\pi}]^2 \,
          [\Lambda + m_{\eta}]^2 \, [ m_{\pi} + m_{\eta} ] } \bigg) \no \\
 & &     - 4 m_K^2 \Big(   b_0 - b_F \Big) 
\eeqa
where we have replaced $\Fnod$ by the pion decay constant $F_{\pi} =
93$ MeV, which is legitimate to the order we are working.
Note that an additional contribution arises for the two kaon nucleon
$\sigma$--terms from the $\pi^0 \eta$ loop.
Furthermore, the $\pi N$ $\sigma$--term is related to the nucleon mass
by the Feynman--Hellman theorem 
$\sigma_{\pi N} (0) = \hat{m} (\partial m_N / \partial \hat{m} )$.

The strange contribution to the nucleon mass is given by the $\sigma$--terms
at zero momentum transfer
\beq \label{eq:sme}
m_s <p|\bar{s}s|p>  =  \bigg( \frac{1}{2}-\frac{m_{\pi}^2}{4 m_K^2}\bigg)  
           \Big( 3 \sigma^{(1)}_{KN}(0) + \sigma^{(2)}_{KN}(0) \Big) 
   \:    +  \:  \bigg( \frac{1}{2} - \frac{m_K^2}{m_{\pi}^2} \bigg)
            \sigma_{\pi N}(0)  \q .
\eeq
This matrix element can  be deduced by means 
of the Feynman--Hellman theorem
$<p|\bar{s}s|p> = \partial m_N / \partial m_s $. 
The strangeness fraction $y$ of the nucleon is given by
\beq	
y = \frac{ 2 <p|\bar{s}s|p> }{ <p | \bar u u + \bar d d|  p>} 
  = \frac{m_{\pi}^2}{\sigma_{\pi N} (0)} \bigg( m_K^2 
    - \frac{1}{2}m_{\pi}^2 \bigg)^{-1} m_s <p|\bar{s}s|p>  
\eeq	
and one defines the quantity $\hat{\sigma}$ via
\beq
\sigma_{\pi N} (0) = \frac{\hat{\sigma}}{1-y} \q .
\eeq

For non--vanishing $t$ we have to distinguish between both regularization
schemes $R_1$ and $R_2$. The general formulae for the scalar form factors
read
\beqa
\sigma_{\pi N} (t)  & = &  \frac{ m_{\pi}^2 }{ 64 \pi F_{\pi}^2 } \,
         \Lambda^4 \, \bigg( 3 [ D+F ]^2  J^{i}(m_{\pi})
        + \frac{1}{3} [ 5D^2-6DF+9F^2 ] J^{i}(m_{K})\no \\
& &     + \frac{1}{9} [ D-3F ]^2   J^{i}(m_{\eta})  \bigg)
        -2 m_{\pi}^2 \Big( b_D + b_F + 2 b_0 \Big) \\
\sigma_{K N}^{(1)} (t)  & = &  \frac{ m_{K}^2 }{ 64 \pi F_{\pi}^2 } \,
         \Lambda^4 \, \bigg( \frac{3}{2} [ D+F ]^2 \, J^{i}(m_{\pi})
         +  [ \frac{7}{3}D^2-2DF+5F^2 ] \, J^{i}(m_{K}) \no \\
& &   + \frac{5}{18} [ D-3F ]^2 \, J^{i}(m_{\eta}) 
            - \frac{1}{6} [D-3F] \, [D+F] \, K^i \bigg) \:
                 - 4  m_K^2 \Big( b_D +  b_0 \Big) \\
\sigma_{K N}^{(2)} (t)  & = &  \frac{ m_{K}^2 }{ 64 \pi F_{\pi}^2 } \,
         \Lambda^4 \, \bigg( \frac{3}{2} [ D+F ]^2 \, J^{i}(m_{\pi})
         + 3 [ D - F ]^2 \, J^{i}(m_{K})
        + \frac{5}{18} [ D-3F ]^2 \, J^{i}(m_{\eta})  \no \\
& &  + \frac{1}{2} [D-3F] \, [D+F] \, K^i \bigg) \: 
                  - 4 m_K^2 \Big(   b_0- b_F  \Big) 
\eeqa
with $i = 1,2$.
Employing the regulator $R_1$ for the evaluation
of the Goldstone boson loops leads to 
\beqa
J^{1}(m_{\phi})  & =& 
    -   \frac{1}{ (m_{\phi}^2 - \Lambda^2)^2} \Bigg( m_{\phi} - \Lambda
  - \frac{ m_{\phi} ( m_{\phi}^2 - \Lambda^2 )}{ (m_{\phi}+\Lambda)^2 -t}\no \\
& &     + \frac{ m_{\phi}^2 - \frac{1}{2} t }{ \sqrt{t} } \bigg[
    \ln \frac{ 2 m_{\phi} + \sqrt{t} }{ 2 m_{\phi} - \sqrt{t} } -
    \ln \frac{ m_{\phi} + \Lambda + \sqrt{t}}{ m_{\phi} + \Lambda-\sqrt{t}}
     \bigg] \, \Bigg) \no \\
K^1   & = &  
    -   \frac{1}{ (m_{\eta}^2 - \Lambda^2)^2} \Bigg( m_{\eta} - \Lambda
  - \frac{ m_{\pi} ( m_{\eta}^2 - \Lambda^2 )}{ (m_{\pi}+\Lambda)^2 -t} \no \\
& &  \qq \qq   + \frac{ m_{\pi}^2 + m_{\eta}^2 - t }{ 2 \sqrt{t} } \bigg[
    \ln \frac{ m_{\pi} + m_{\eta} + \sqrt{t} }{ m_{\pi} 
                              + m_{\eta} - \sqrt{t} } -
    \ln \frac{ m_{\pi} + \Lambda + \sqrt{t}}{ m_{\pi} + \Lambda-\sqrt{t}}
     \bigg] \, \Bigg) \no \\
& &  -   \frac{1}{ (m_{\pi}^2 - \Lambda^2)^2} \Bigg( m_{\pi} - \Lambda
  - \frac{ m_{\eta} ( m_{\pi}^2 - \Lambda^2 )}{ (m_{\eta}+\Lambda)^2 -t} \no \\
& &  \qq \qq   + \frac{ m_{\eta}^2 + m_{\pi}^2 - t }{ 2 \sqrt{t} } \bigg[
    \ln \frac{ m_{\pi} + m_{\eta} + \sqrt{t} }{ m_{\pi} 
                              + m_{\eta} - \sqrt{t} } -
    \ln \frac{ m_{\eta} + \Lambda + \sqrt{t}}{ m_{\eta} + \Lambda-\sqrt{t}}
     \bigg] \, \Bigg) \q ,
\eeqa
whereas for the modified regulator $R_2$ one obtains
\beqa
J^{2}(m_{\phi})  & =& 
    -   \frac{1}{ (m_{\phi}^2 - \Lambda^2)^2} \, 
      \frac{1}{\sqrt{t}} \bigg( \,\,   [ m_{\phi}^2 -
\frac{1}{2} t ] \, \ln \frac{ 2 m_{\phi} +\sqrt{t} }{ 2 m_{\phi} 
- \sqrt{t}}\no\\
&&     + [ \Lambda^2 - \frac{1}{2} t] \, \ln \frac{ 2 \Lambda 
     +\sqrt{t} }{ 2 \Lambda - \sqrt{t}} \, - \, [ m_{\phi}^2 + \Lambda^2 - t ]
\, \ln \frac{ m_{\phi} + \Lambda + \sqrt{t}}{ m_{\phi} + \Lambda-\sqrt{t}}
     \,\,  \bigg) \no \\
K^2   & = &  
   -   \frac{1}{ (m_{\eta}^2 - \Lambda^2)\,  (m_{\pi}^2 - \Lambda^2)}
    \frac{1}{\sqrt{t}} \bigg(\, \,   [m_{\pi}^2 + m_{\eta}^2 - t]
  \,   \ln \frac{ m_{\pi} + m_{\eta} + \sqrt{t} }{ m_{\pi} 
                              + m_{\eta} - \sqrt{t} }   \no \\
&&  - [m_{\pi}^2 + \Lambda^2 - t] \, 
   \ln \frac{ m_{\pi} + \Lambda + \sqrt{t}}{ m_{\pi} + \Lambda-\sqrt{t}}
- \, [m_{\eta}^2 + \Lambda^2 - t] \, 
 \ln \frac{ m_{\eta} + \Lambda + \sqrt{t}}{ m_{\eta} + \Lambda-\sqrt{t}}\no \\
&&+ [ 2 \Lambda^2 - t] \ln \frac{2 \Lambda 
+ \sqrt{t}}{ 2 \Lambda-\sqrt{t}} \,\,   \bigg) \q .
\eeqa
One is in particular interested in the shifts of the $\sigma$--terms 
to the Cheng--Dashen points. These are $t= 2 m_{\pi}^2$ and $t= 2 m_K^2$
for the $\pi N$ and $K N$ $\sigma$--terms, respectively. The 
$\sigma$--terms can acquire imaginary parts depending on the values
for $t$ and $\Lambda$. Since we will consider $\Lambda$ only in the 
range from 300 MeV to 600 Mev the shift of the $\pi N$ $\sigma$--term
is real. On the other hand, the shifts of the 
two $KN$ $\sigma$--terms acquire an
imaginary part at $t = 4 m_{\pi}^2$ and also at $t = ( m_{\pi} + \Lambda)^2$
for sufficiently small values of $\Lambda$ in both regularization schemes.
In the regularization scheme $R_2$ there is an additional branch cut
starting at $t = 4 \Lambda^2$. For increasing $\Lambda$ only the
first branch cut $t= 4 m_{\pi}^2$ from the pion loop remains. 
This agrees with
the dimensional regularization scheme which is recovered for $\Lambda
\rightarrow \infty$.
Since we choose the Gell-Mann--Okubo value for the $\eta$ mass, the $\pi
\eta$ loop does not acquire an imaginary part below $t= 2 m_K^2 <
( m_{\pi} + m_{\eta} )^2 $. For the physical mass of the $\eta$
this contribution is tiny compared to the other parts.
Before presenting the numerical results we will include the decuplet
in the next section.

\section{Inclusion of the decuplet}
In general it is assumed that baryon resonance states are much heavier 
compared to the lowest--lying baryon octet. In this case they can be
integrated out and replaced by counterterms that do not include these
resonance states explicitely. However, while this might be a reasonable
procedure for heavier resonances like the Roper--octet, it is a
questionable assumption for the decuplet.
The low--lying decuplet is separated from the octet by only $\Delta = 231$
MeV in average which is much smaller than the $K$ or the $\eta$ mass. 
Furthermore, the $\Delta(1232)$ couples strongly to the $\pi N$ sector
and its contribution plays an important role in the channels wherein
this effect is possible.
In the meson sector, the first resonance is the vector meson $\rho$ with
a mass of 770 MeV which is considerably heavier than the
Goldstone bosons.
It was therefore argued in \cite{JM} to include the spin--3/2 decuplet
as explicit degrees of freedom.
In the framework of conventional heavy baryon CHPT it was found
that intermediate $\Delta(1232)$ states give a contribution of 7.5 MeV to the
$\pi N$ $\sigma$--term shift which is as large as the contribution from the
the octet alone \cite{BKM}.

In this section we will include the decuplet fields and the resulting
loop integrals are evaluated by using the regulator $R_2$.
A similar analysis with the regulator $R_1$ instead is not possible
since then divergent integrals arise for non--vanishing momentum $t$.
The pertinent interaction Lagrangian between the spin--3/2 fields
-- denoted by the Rarita--Schwinger fields $T_{\mu}$ --, the baryon octet
and the Goldstone bosons reads
\beq
{\cal L}_{\phi B T} = - i \, \, \bar{T}^{\mu} \, v \cdot D \, T_{\mu} \:
          + \Delta \,   \bar{T}^{\mu} \,   T_{\mu} \:           \:
          + \frac{C}{2} \, \Big( \,\bar{T}^{\mu} \, u_{\mu} B \, 
                     + \, \bar{B} \, u_{\mu}\, T^{\mu}\, \Big) 
\eeq
where we have suppressed the flavor $SU(3)$ indices. In the heavy mass
formulation the fields $T_{\mu}$ satisfy the condition $ v \cdot T = 0$. 
The coupling constant $C = 1.2 ... 1.8$ can be determined from the
strong decays $T \rightarrow B \pi$.
After integrating out the heavy degrees of freedom from the relativistic
Lagrangian there is still a remaining mass dependence which is proportional
to the average octet--decuplet splitting $\Delta$ and does not vanish
in the chiral limit.
In the Feynman rules the mass splitting $\Delta$ is contained in the decuplet
propagator
\beq
\frac{i}{ v \cdot l \, \, - \Delta + \, \, i \epsilon}      
\, \, \bigg( v_{\mu} v_{\nu} - g_{\mu \nu} -\frac{4}{3}S_{\mu}S_{\nu} \bigg)
\eeq
The appearance of the mass scale $\Delta$ destroys in the case of
dimensional regularization the one--to--one correspondence between
meson loops and the expansion in small momenta and quark masses.
No further complications arise in our case since the strict chiral
counting scheme has already been spoilt by introducing the scale $\Lambda$.

For zero momentum transfer $t=0$ the decuplet contributions
to the $\sigma$--terms read
\beqa
\delta \sigma_{\pi N} (0)  & = &
        - \frac{ m_{\pi}^2 C^2}{ 96 \pi F_{\pi}^2 } \,
         \Lambda^4 \, \bigg( 8 H (m_{\pi}) + H (m_{K}) \bigg) \\
\delta \sigma_{K N}^{(1)} (0)  & = &  
         - \frac{ m_{K}^2 C^2}{96 \pi F_{\pi}^2 } \,
         \Lambda^4 \, \bigg( 4 H (m_{\pi}) + \frac{4}{3} H (m_{K}) \bigg) \\
\delta \sigma_{K N}^{(2)} (0)  & = & 
        -  \frac{ m_{K}^2 C^2}{96 \pi F_{\pi}^2 } \,
         \Lambda^4 \, \bigg( 4 H (m_{\pi}) + 2 H (m_{K}) \bigg)
\eeqa
with
\beqa
H (m_{\phi})   & = &
    \frac{1}{ (\Lambda^2 - m_{\phi}^2)^3} \, \bigg( - \Delta
    [\Lambda^2 - m_{\phi}^2] + \, \frac{1}{2} \Delta \, [ 4 \Delta^2 
    - 3 \Lambda^2 - 3 m_{\phi}^2 ]\, \ln \frac{m_{\phi}^2}{\Lambda^2} \no \\
& & \qq \qq
  +\, [ 4 \Delta^2 - 3 \Lambda^2 - m_{\phi}^2 ]\, \sqrt{\Delta^2 - m_{\phi}^2}
    \, \ln \bigg[ \frac{\Delta}{m_{\phi}}   
      + \sqrt{\frac{\Delta^2}{m_{\phi}^2}-1} \: \bigg]\no \\
& & \qq \qq   
    +\, [ 4 \Delta^2 - \Lambda^2 - 3 m_{\phi}^2 ]
   \, \sqrt{\Lambda^2 - \Delta^2}\, \arccos \frac{\Delta}{\Lambda}\bigg) \: ;
      \; \;  \; \; \; \; \mbox{for}  \; \; m_{\phi} < \Delta  \no \\
H (m_{\phi})   & = &
    \frac{1}{ (\Lambda^2 - m_{\phi}^2)^3} \, \bigg( - \Delta
    [\Lambda^2 - m_{\phi}^2] + \, \frac{1}{2} \Delta [ 4 \Delta^2 
    - 3 \Lambda^2 - 3 m_{\phi}^2 ] \ln \frac{m_{\phi}^2}{\Lambda^2} \no \\
& & \qq \qq - [ 4 \Delta^2 - 3 \Lambda^2 
              - m_{\phi}^2 ] \sqrt{m_{\phi}^2 - \Delta^2}
    \arccos \frac{\Delta}{m_{\phi}} \no \\
& & \qq \qq    + [ 4 \Delta^2 - \Lambda^2 - 3 m_{\phi}^2 ]
    \sqrt{\Lambda^2 - \Delta^2} \arccos \frac{\Delta}{\Lambda} \: \bigg) \: ;
      \; \; \; \; \; \;  \mbox{for}  \; \;   m_{\phi} > \Delta  
\eeqa
where we required $\Lambda > \Delta$. For the limit
$ \Lambda >> m_{\phi} $ we recover up to some constant terms the result
from dimensional regularization \cite{BKM}
\beq
H (m_{\phi})  \stackrel{m_{\phi}<<\Lambda}{\longrightarrow}
- \frac{3}{2} \bigg( \, \Delta \Big[ \ln \frac{m_{\phi}^2}{\Lambda^2}
   + \frac{2}{3} \Big] - 2 \sqrt{m_{\phi}^2 - \Delta^2} 
    \arccos \frac{\Delta}{m_{\phi}}+ 
   \frac{\pi}{3} \Lambda + \ldots \bigg)
\eeq
for the case $m_{\phi} > \Delta$ and an analogous result 
for $ m_{\phi}< \Delta$.
The constant terms can again be absorbed into a renormalization
of the parameters $ b_{0,D,F}$.
The results for the scalar form factors for general $t$ can be found
in App.~A.

\section{Results and discussion}
In this section we present the numerical results for the calculation
of the $\sigma$--terms.
We consider first the case with no resonances.
The values for our parameters are $D=0.75$, $F=0.5$, $F_{\pi}=93$ MeV,
$m_{\pi}=138$ MeV, $m_K=495$ MeV, and for the mass of the $\eta$ 
we use the GMO value for
the pseudoscalar mesons $m_{\eta}=566$ MeV.
The differences for $F_{\pi}$ and $m_{\eta}$ to  $\Fnod$--the pseudoscalar
decay constant in the chiral limit-- and to the physical mass of $\eta$,
respectively, appear only at higher orders.
We will restrict ourselves to these central values of the parameters
since a small variation in these parameters does only lead to some minor
changes in the results.

In baryon chiral perturbation theory, the transition between short and
long distance occurs around a distance scale of $\sim$1 fermi, or a momentum
scale of $\sim$200 MeV. This corresponds to the measured size of a baryon.
The effective field
theory treats the baryons and pions as point particles. This is
appropriate for the very long distance physics. 
However, for propagation at distances less then the separation
scale, the point particle theory is not an accurate representation of
the physics. The composite substructure becomes manifest below this
point.

Of course, the cutoff $\Lambda$ should not be taken so low in energy that it
removes any truly long distance physics. Also, while it can in principle be
taken much larger than the separation scale, this will lead to the
inclusion of spurious short distance physics which can upset the
convergence of the expansion. It is ideal to take the cutoff slightly
above the separation scale so that all of the long distance physics,
but little of the short distance physics, is included.
Therefore, we will vary the cutoff in the range
$\Lambda\geq 1/<r_B>\sim 300-600$ MeV.

The four unknown parameters $ b_{0,D,F}$ and $\mnod$ have to
be fixed from phenomenology. We will choose the four different baryon masses
in the isospin limit and the value of $\sigma_{\pi N} \simeq 45$ MeV 
to perform a least--squares fit for these parameters.
The explicit calculation of the masses in the cutoff regularization scheme
to the order we are working can be found in \cite{DHB}.
As outlined in this work the asymptotic mass--independent component
of the mass integral which is proportional to $\Lambda^3$ is removed
by redefining $\mnod$
\beq
{\mnod}_r = \mnod - (5 D^2 + 9 F^2) \frac{\Lambda^3}{48 \pi F_{\pi}^2} \q .
\eeq
In Tables~1 and ~3 we present the renormalized values $\mnod_r$
after absorbing the consatnt pieces from the mass integral.
We are then able to make predictions for the $KN$ $\sigma$--terms
and for the shifts to the Cheng--Dashen points.
The results of this calculation are shown in Tab.~1 and  Tab.~2. 
In the first Table, 
we present besides the values for $ b_{0,D,F}$ and $\mnod$
the two $KN$ $\sigma$--terms at zero momentum transfer, the strange quark
matrix element from Eq.~(\ref{eq:sme}),  the strange quark fraction $y$,
$\hat{\sigma}$ and the deviation from
the Gell-Mann--Okubo relation $[3M_\Lambda+M_\Sigma-2M_N-2M_\Xi]/4$ 
between the  baryon masses which is experimentally about 6.5 MeV.
Within the accuracy of the calculation, the $KN$ $\sigma$--terms
turn out to be
\beqa
\sigma_{KN}^{(1)}  (0)   &=&  400  \pm  30 \; \mbox{MeV} \no \\
\sigma_{KN}^{(2)}  (0)   &=&  270  \pm  20 \; \mbox{MeV} 
\eeqa
where the uncertainty stems from the variation in the cutoff $\Lambda$.
For the other quantities we have
\beq
m_s <p|\bar{s}s|p> = 150 \pm 50 \; \mbox{MeV} \: , 
\qq y = 0.25 \pm  0.05 \: , \qq 
\hat{\sigma} = 33 \pm 3 \; \mbox{MeV} \q .
\eeq
The value for $y$ is within the band deduced in \cite{GLS},
$y = 0.15 \pm 0.10$, and the value for $\hat{\sigma}$ agrees nicely
with Gasser's estimate $\hat{\sigma} = 33 \pm 5$ MeV, given in \cite{G}.
Therein, the author comes to the conclusion that the lowest non--analytic
corrections to the baryon masses and the $\pi N$ $\sigma$--term are so
large, that chiral perturbation theory is meaningless in that case.
He proposes a meson--cloud model by introducing a cutoff which regularizes
the divergent integrals. The cutoff is provided by the square of the
axial vector form factor which enters the expression of the propagator
and is similar to $R_1^2$ in our notation.
The deviation from the Gell-Mann-Okubo relation due to loops is found to
be quite small in dimensional regularization, 
primarily due to the (accidental) feature that it is
proportional to $D^2-3F^2  <<1$.
We, therefore, expect this deviation to be even smaller in the
cutoff scheme.
The chiral expansions of the $\sigma$--terms in dimensional
regularization read
\beqa
\sigma_{\pi N} (0)      &=&   82.7 - 37.7  \: \mbox{MeV}\q = 
\q 45.0  \: \mbox{MeV}   \no \\
\sigma_{K N}^{(1)} (0)  &=&   763.0 - 605.6 \:  \mbox{MeV} \q = 
 \q 157.4 \: \mbox{MeV}  \no \\
\sigma_{K N}^{(2)} (0)  &=&   177.1 - 56.6  \: \mbox{MeV}   \q = 
 \q 120.5\: \mbox{MeV} \q . 
\eeqa
The chiral expansions of the $\sigma$--terms 
in the cutoff scheme for $\Lambda = 400$ MeV are
\beqa
\sigma_{\pi N} (0)      &=&   36.2 + 8.8  \: \mbox{MeV}  \q = 
\q 45.0  \: \mbox{MeV} \no \\
\sigma_{K N}^{(1)} (0)  &=&   340.5 + 66.6 \: \mbox{MeV}  \q = 
\q 407.1 \: \mbox{MeV} \no \\
\sigma_{K N}^{(2)} (0)  &=&   227.0 + 48.1 \: \mbox{MeV}  \q = 
\q 275.1 \: \mbox{MeV}
\eeqa
where the first number denotes the lowest order contribution from the
tree level result and the second number the Goldstone boson loop
contribution.
The chiral expansions in the cutoff scheme are much improved with respect
to the case of dimensional regularization, especially for
$\sigma_{\pi N}$ and $\sigma_{KN}^{(1)}$.
While in dimensional regularization the contributions of two
successive chiral orders are of opposite sign and tend to cancel
each other -- a common feature in this regularization scheme --,
there is a clear convergence in the cutoff scheme. 
We also examined our results by varying the value of the $\pi N$
$\sigma$--term by $\pm 10$ MeV. This change alters the value of
$b_0$ from $-0.29$ to $- 0.55$ GeV$^{-1}$ which has a quite dramatic
impact on the $KN$ $\sigma$--terms and the value of $m_s <p|\bar{s}s|p>$.
This is in agreement with the calculation in dimensional regularization
\cite{BKM}. We will restrict ourselves to 
the central value of $\sigma_{\pi N} = 45$ MeV
\cite{GLS} in our analysis.
The changes in a calculation with the physical mass of the $\eta$, 
$m_{\eta} = 549$ MeV, are negligible.

In Table~2 we list the shifts of the scalar form factors to the
Cheng--Dashen points. We are able to compare both
regularization schemes $R_1$ and $R_2$ since these differ only for
non--vanishing off--shell momenta of the baryons.
While there is agreement for the $\pi N$ $\sigma$--term, both
regularization schemes differ considerably in the $KN$ $\sigma$--terms
which depend strongly on the cutoff $\Lambda$. This might indicate
that for these quantities higher orders play an essential role.
Clearly a definite statement about the shifts to the Cheng--Dashen
points for the $KN$ $\sigma$--terms cannot be made.
The $\pi N$ $\sigma$--term shift agrees in both regularization schemes
and we find
\beq  \label{shift}
\sigma_{\pi N} ( 2 m_{\pi}^2) -\sigma_{\pi N} (0) = 4 \pm 1 \; \mbox{MeV}\q .
\eeq

This value is in agreement with the result
$\Delta \sigma_{\pi N} = 5 \pm 1$ MeV of the complete fourth order
calculation in conventional heavy baryon chiral perturbation theory \cite{B}.
On the other hand, it is smaller than the empirical value found in
\cite{GLS}. The main contribution to the $\pi N$ $\sigma$--term shift
in the dispersive calculation in that paper comes from an energy region
in which the one--loop approximation is off by a factor of two.
Therefore, we expect contributions from higher chiral orders to be 
significant. It remains to be seen how higher order corrections not yet
accounted for will modify Eq.~(\ref{shift}).

Adding the decuplet, we set $\Delta=231$ MeV, which is the average 
octet--decuplet mass splitting, and the value of the coupling constant $C$ 
is given by $C=1.5$ from an overall fit to the decuplet decays \cite{JM2}.
To perform a least--squares fit for the parameters $b_{0,D,F}$ and $\mnod$
we have to include the decuplet contributions to the octet baryon masses,
see App.~B.
The results of the fit can be found in Tab.~3 and  Tab.~4. 
We absorbed again the asymptotic mass--independent 
component of the mass integral
by redefining $\mnod_r$ 
\beq
\mnod_r \q \rightarrow \q \mnod_r  - 
      \frac{5 C^2 \Lambda^2}{ 24 \pi^2 F_{\pi}^2}
         \Big( -\frac{\pi}{4} \Lambda + \frac{1}{2}\Delta \Big) \q .
\eeq
It turns out that there are no significant changes in the results
as in the case of dimensional regularization.
We obtain for the $KN$ $\sigma$--terms
\beqa
\sigma_{KN}^{(1)}  (0)   &=&  380  \pm  40 \; \mbox{MeV} \q , \no \\
\sigma_{KN}^{(2)}  (0)   &=&  250  \pm  30 \; \mbox{MeV} \q .
\eeqa
For the strange quark contribution to the nucleon we have
\beq
m_s <p|\bar{s}s|p> = 110 \pm 60 \; \mbox{MeV} \: , 
\qq y = 0.20 \pm  0.12 \: , \qq  
\hat{\sigma} = 35 \pm 6 \; \mbox{MeV} \q.
\eeq
Note the large value of the matrix element
$m_s <p|\bar{s}s|p> = -671$ MeV in dimensional regularization.
This time the chiral expansions of the $\sigma$--terms 
in the cutoff scheme for $\Lambda = 400$ MeV are
\beqa
\sigma_{\pi N} (0)      &=&   31.2 + 13.8  \: \mbox{MeV}  \q = 
\q 45.0  \: \mbox{MeV} \no \\
\sigma_{K N}^{(1)} (0)  &=&   289.7 + 100.4 \: \mbox{MeV}  \q = 
\q 390.1 \: \mbox{MeV} \no \\
\sigma_{K N}^{(2)} (0)  &=&   173.9 + 83.4 \: \mbox{MeV}  \q = 
\q 257.3 \: \mbox{MeV} \q ,
\eeqa
whereas the chiral expansions of the $\sigma$--terms in dimensional
regularization read
\beqa
\sigma_{\pi N} (0)      &=&   140.7 - 95.7  \: \mbox{MeV}\q = 
\q 45.0  \: \mbox{MeV}   \no \\
\sigma_{K N}^{(1)} (0)  &=&   1060 - 1106 \:  \mbox{MeV} \q = 
 \q -46 \: \mbox{MeV}  \no \\
\sigma_{K N}^{(2)} (0)  &=&   557 - 658  \: \mbox{MeV}   \q = 
 \q -101 \: \mbox{MeV} \q .
\eeqa
As in the case without resonances the convergence of the chiral
series in the cutoff scheme is significantly improved with respect to 
dimensional regularization.
We do not observe a dramatic change in the results as
suggested by employing dimensional regularization.
For the evaluation
of the shifts of the scalar formfactors to the Cheng--Dashen points
we applied the regularization scheme $R_2$ since the scheme $R_1$ leads to
divergent integrals.
For the $\sigma_{\pi N}$ shift we find
\beq
\sigma_{\pi N} ( 2 m_{\pi}^2) -\sigma_{\pi N} (0) = 6 \pm 1 \; \mbox{MeV}\q .
\eeq
There is still a sizeable uncertainty in the $KN$ $\sigma$--terms, and
we present only the results for $\Lambda = 400$ and $500$ MeV in Table~4.

\section{Summary}
In this paper, we have evaluated the $\pi N$ and $KN$ $\sigma$--terms  
and scalar form factors by using a cutoff regularization.

\begin{enumerate}

\item[$\circ$] First, we calculated the $\sigma$--terms by using
the next--to--leading order Lagrangian for the Goldstone bosons and the 
lowest--lying baryon octet in the heavy baryon formulation.
The Goldstone boson integrals are evaluated by using a simple
dipole regulator  with a cutoff $\Lambda$
proposed in \cite{DHB}. We also used a modified
regulator which is similar to the first one for vanishing off--shell
momenta of the external baryons. We have given the complete expressions
for the $\sigma$--terms up to the order $q^3$, where $q$ is an external
momentum or meson mass.
The cutoff parameter induces an additional mass scale that does not vanish
in the chiral limit and, therefore, destroys the strict chiral counting
scheme. We are able to show that to the order we are working the
physics does not depend on $\Lambda$, since one is able to absorb
the effects of $\Lambda$ into a renormalization of the coupling
constants.

\item[$\circ$] The spin--3/2 decuplet is separated from the octet
by 231 MeV in average which is smaller than the kaon or eta mass.
Therefore, we proceeded by adding the decuplet to the effective theory.
Performing the calculation with the regulator from \cite{DHB} 
leads to divergent integrals for non--vanishing off-shell momenta. 
One obtains finite results for the other regulator.

\item[$\circ$] There are four unknown parameters in the theory --
the coupling constants $b_{0,D,F}$ from the Lagrangian of chiral order
$q^2$ and the baryon mass in the chiral limit $\mnod$ -- which have to
be fixed from phenomenology. We choose the four baryon masses
in the isospin limit $(N,\Sigma,\Lambda,\Xi)$ and the value
$\sigma_{\pi N} (0) = 45$ MeV to perform a least--squares fit
for these parameters. 
In our analysis the cutoff parameter ranges from 300 to 600 MeV
to account for all the long distance physics, but little of the
short distance physics, which are not described appropriately by
the effective theory, is included.
Predictions for the $KN$ $\sigma$--terms and the strange  contribution
to the nucleon mass are made.
The results without the decuplet are $\sigma_{KN}^{(1)}(0) = 400 \pm 30$
MeV and $\sigma_{KN}^{(2)}(0) = 270 \pm 20$ MeV for the two
$KN$ $\sigma$--terms (accounting for the uncertainty in $\Lambda$). 
The strange contribution to the nucleon mass
is $<p|m_s \bar{s}s|p> = 150 \pm 50$ MeV which translates into the
strangeness fraction $y = 0.25 \pm 0.05$ and $\hat{\sigma} = 33 \pm 3$
MeV. The results are in good agreement with previous calculations 
\cite{GLS,G}.
While a definite statement about the convergence of the chiral
expansions for the $\sigma$--terms cannot be made in dimensional 
regularization, there is a clear convergence in the cutoff scheme.
The $\pi N$ $\sigma$--term shift to the Cheng--Dahen point is $4 \pm 1$ MeV
in both cutoff schemes. 
This number is in agreement with the complete fourth order calculation in
conventional heavy baryon chiral perturbation theory. But this value
is smaller than the dispersive calculation of \cite{GLS}.
It remains to be seen how higher order corrections not yet accounted for 
will modify this result.
The shifts for the $KN$ $\sigma$--terms
depend strongly on the value of $\Lambda$ which might indicate
that higher chiral orders are important.
In order to include the decuplet states into the fit, one has to
account for the decuplet contributions to the baryon masses. Again a
least--squares fit is performed and it turns out that there are no 
significant changes in the results.
One obtains
$\sigma_{KN}^{(1)}(0) = 380 \pm 40$ MeV, 
$\sigma_{KN}^{(2)}(0) = 250 \pm 30$ MeV,
$<p|m_s \bar{s}s|p> = 110 \pm 60$ MeV, $y= 0.20 \pm 0.12$
and $\hat{\sigma} = 35 \pm 6$ MeV.
For the $\pi N$ $\sigma$--term shift we obtain $6 \pm 1$ MeV.

\end{enumerate}

\section*{Acknowledgements}
I am grateful to John Donoghue for useful discussions and
reading the manuscript.
This work was supported in part by the Deutsche Forschungsgemeinschaft.

\appendix 
\def\theequation{\Alph{section}.\arabic{equation}}
\setcounter{equation}{0}
\section{Decuplet contributions to the scalar formfactors} \label{app.a}
The decuplet contributions to the scalar formfactors for non--vanishing
momentum transfer $t$ can be presented
as follows
\beqa
\delta \sigma_{\pi N} (t)  & = &
        - \frac{ m_{\pi}^2 C^2}{ 96 \pi F_{\pi}^2 } \,
      \Lambda^4 \, \bigg( 8 \tilde{H} (m_{\pi}) + \tilde{H} (m_{K}) \bigg) \\
\delta \sigma_{K N}^{(1)} (t)  & = &  
         - \frac{ m_{K}^2 C^2}{96 \pi F_{\pi}^2 } \,
   \Lambda^4 \, \bigg( 4 \tilde{H} (m_{\pi}) 
    + \frac{4}{3} \tilde{H} (m_{K}) \bigg) \\
\delta \sigma_{K N}^{(2)} (t)  & = & 
        -  \frac{ m_{K}^2 C^2}{96 \pi F_{\pi}^2 } \,
         \Lambda^4 \, \bigg( 4 \tilde{H} (m_{\pi}) + 2 \tilde{H} (m_{K}) \bigg)
\eeqa
with
\beqa
\tilde{H} (m_{\phi})  & = &
      \frac{1}{ (\Lambda^2 - m_{\phi}^2)^2} \, \Bigg( -2 \Delta
  \bigg[ \sqrt{ \frac{m_{\phi}^2}{t} - \frac{1}{4}} 
                       \arcsin \frac{\sqrt{t}}{2 m_{\phi}}
  +      \sqrt{ \frac{\Lambda^2}{t} - \frac{1}{4}} 
                       \arcsin \frac{\sqrt{t}}{2 \Lambda}  \bigg] \no \\
&& + \Delta \bigg[ \frac{m_{\phi}^2-\Lambda^2}{2t} 
       \ln \frac{m_{\phi}^2}{\Lambda^2}  +
 \frac{1}{t} \beta \Big( \mbox{artanh} \frac{m_{\phi}^2-\Lambda^2 +t}{\beta}
 - \mbox{artanh} \frac{m_{\phi}^2-\Lambda^2 -t}{\beta}\Big) \bigg] \no \\
&& - [ \Delta^2 - \Lambda^2 + \frac{1}{2} t ] \int_0^1 dx \,
     \frac{1}{\sqrt{\Lambda^2 - \Delta^2 - x(1-x)t}}
     \arccos \frac{\Delta}{\sqrt{ \Lambda^2- x(1-x)t}}\no \\
&& + [ 2 \Delta^2 - m_{\phi}^2 - \Lambda^2 + t ] \int_0^1 dx \,
     \frac{1}{\sqrt{\Lambda^2 - x [\Lambda^2-m_{\phi}^2]- \Delta^2 - x(1-x)t}}
     \no \\
&&  \times
   \arccos \frac{\Delta}{\sqrt{ \Lambda^2-x[\Lambda^2-m_{\phi}^2]- x(1-x)t}}
  - [ \Delta^2 - m_{\phi}^2 + \frac{1}{2} t ] \, f (m_{\phi}) 
 \;\Bigg) 
\eeqa
with
\beq
\beta = \sqrt{ ( m_{\phi}^2 - \Lambda^2 -t )^2 -4 t \Lambda^2}
\eeq
and
\beqa
f (m_{\phi}) &=& \int_0^1 dx \, 
     \frac{1}{\sqrt{\Delta^2 - m_{\phi}^2 + x(1-x)t}}
     \ln \frac{\Delta + \sqrt{\Delta^2 - m_{\phi}^2 + x(1-x)t}}{
       \sqrt{m_{\phi}^2 - x(1-x)t}} \; ;
    \; \; \; \; \; \mbox{for}  \; \; m_{\phi} < \Delta  \no \\
f (m_{\phi}) &=& \int_0^1 dx \, 
     \frac{1}{\sqrt{m_{\phi}^2 - \Delta^2  - x(1-x)t}}
      \arccos \frac{\Delta}{\sqrt{m_{\phi}^2 - x(1-x)t}} \; ;
      \; \; \; \; \; \mbox{for}  \; \; m_{\phi} > \Delta  
\eeqa
where we required $\Lambda > \Delta$.
We presented the result for sufficiently small $t$. With increasing $t$ 
the squareroots 
become imaginary and one has to continue $\tilde{H} (m_{\phi})$ analytically.

\def\theequation{\Alph{section}.\arabic{equation}}
\setcounter{section}{1}
\setcounter{equation}{0}
\section{Decuplet contributions to the masses} \label{app.b}
In this Appendix we give the results for the decuplet contributions to the
masses.
They can be written in the form
\beq
\delta M_B = \frac{ C^2}{ 24 \pi F_{\pi}^2 } \,\Lambda^4 \, \bigg( 
  \alpha^{\pi}_{B} M(m_{\pi}) +  \alpha^{K}_{B} M(m_{K}) + 
  \alpha^{\eta}_{B} M(m_{\eta})  \bigg)
\eeq
with coefficients
\beqa
&&\alpha^{\pi}_{N} = 4\, ,\:\;  \alpha^{K}_{N} = 1 \, ,\: \; 
\alpha^{\eta}_{ N} = 0\, ; \:\;
\alpha^{\pi}_{\Sigma} = \frac{2}{3} \, ,\;\:
\alpha^{K}_{\Sigma} = \frac{10}{3} \, ,\;\:  
\alpha^{\eta}_{ \Sigma} = 1\, ;\no \\ 
&&\alpha^{\pi}_{\Lambda} = 3\, ,\;\: \alpha^{K}_{\Lambda} = 2 \, ,\; \:
\alpha^{\eta}_{ \Lambda} = 0\, ;\:\;
\alpha^{\pi}_{\Xi} = 1\, ,\; \: \alpha^{K}_{\Xi} = 3 \, ,\; \:
\alpha^{\eta}_{\Xi} = 1\, ;
\eeqa
and
\beqa
M(m_{\phi})  & = & \frac{1}{ (\Lambda^2 - m_{\phi}^2)^2} \, \bigg( \,
     \frac{1}{2} \Delta [ \frac{3}{2}m_{\phi}^2
     - \Delta^2]  \ln \frac{m_{\phi}^2}{\Lambda^2}
    - \frac{1}{4} \Delta [ m_{\phi}^2 - \Lambda^2 ] \no \\
&& \qq \qq - [\Delta^2 - m_{\phi}^2]^{3/2} 
    \ln \bigg[ \frac{\Delta}{m_{\phi}}   
      + \sqrt{\frac{\Delta^2}{m_{\phi}^2}-1} \bigg] +
     [\Lambda^2 - \Delta^2]^{3/2} \arccos \frac{\Delta}{\Lambda} \no \\
&& \qq \qq +\frac{3}{2} [ m_{\phi}^2 - \Lambda^2 ][\Lambda^2 - \Delta^2]^{1/2}
    \arccos \frac{\Delta}{\Lambda}  \: \bigg)  \: ;
      \; \; \; \; \; \; \; \; \mbox{for}  \; \; m_{\phi} < \Delta  \no \\
M(m_{\phi})  & = & \frac{1}{ (\Lambda^2 - m_{\phi}^2)^2} \, \bigg(\, 
     \frac{1}{2} \Delta [ \frac{3}{2}m_{\phi}^2
     - \Delta^2]  \ln \frac{m_{\phi}^2}{\Lambda^2}
    - \frac{1}{4} \Delta [ m_{\phi}^2 - \Lambda^2 ] \no \\
&& \qq \qq - [m_{\phi}^2 - \Delta^2]^{3/2} 
    \arccos \frac{\Delta}{m_{\phi}} +
     [\Lambda^2 - \Delta^2]^{3/2} \arccos \frac{\Delta}{\Lambda} \no \\
&& \qq \qq +\frac{3}{2} [ m_{\phi}^2 - \Lambda^2 ][\Lambda^2 - \Delta^2]^{1/2}
    \arccos \frac{\Delta}{\Lambda}  \:  \bigg)  \: ;
      \; \;  \; \; \; \; \; \;\mbox{for}  \; \; m_{\phi} > \Delta  
\eeqa
where we required $\Lambda > \Delta$.

\newpage

\section*{Table captions}

\begin{enumerate}

\item[Table 1] Given are the LECs $ b_{0,D,F}$, the renormalized baryon mass
               in the chiral limit $\mnod_r$, the two $KN$ $\sigma$--terms, 
               the strange quark matrix element (SME),  the strange quark 
               fraction $y$, $\hat{\sigma}$ 
               and the deviation from the Gell-Mann--Okubo 
               relation  $[3M_\Lambda+M_\Sigma-2M_N-2M_\Xi]/4$  
               in dimensional regularization and for various values of the
               cutoff parameter $\Lambda$ in MeV.

\item[Table 2] Shifts to the Cheng--Dashen points in MeV in 
               dimensional regularization and for both
               regularization schemes $R_1$ and $R_2$. 
               The cutoff parameter $\Lambda$ is given in MeV.

\item[Table 3] Results of the calculation including the spin--3/2
               decuplet. The cutoff parameter $\Lambda$ is given in MeV.

\item[Table 4] Shifts to the Cheng--Dashen points in MeV in 
               dimensional regularization and for the
               regularization scheme $R_2$ and including the decuplet.
               The cutoff parameter $\Lambda$ is given in MeV.

\end{enumerate}

\newpage

\begin{center}

\begin{table}[bht]  \label{tab1}
\begin{center}
\begin{tabular}{l|c|c|c|c|c}
  & dim. & $\Lambda=300$&$\Lambda=400$&$\Lambda=500$&$\Lambda=600$\\
\hline
$b_D \;\;$ [GeV$^{-1}$]             & 0.008 & 0.068 & 0.070 & 0.072 & 0.075 \\
$b_F \;\; $ [GeV$^{-1}$]            &-0.606 &-0.197 &-0.185 &-0.171 &-0.153 \\
$b_0 \;\; \:$ [GeV$^{-1}$]          &-0.786 &-0.459 &-0.417 &-0.371 &-0.320 \\
$\mnod_r \; \; $[GeV]           & 1.011 & 0.701 & 0.720 & 0.738 & 0.755 \\
$\sigma_{KN}^{(1)} (0)\; $ [MeV]  & 157.4 & 420.9 & 407.1 & 393.5 & 380.6 \\
$\sigma_{KN}^{(2)} (0)\; $ [MeV]  & 120.4 & 285.5 & 275.0 & 265.0 & 255.8 \\
SME $\; \; \;\:$ [MeV]            &-271.6 & 187.5 & 162.6 & 138.2 & 115.2 \\
$y$                               &-0.488 & 0.337 & 0.292 & 0.248 & 0.207 \\
$\hat{\sigma} \qq \q  $ [MeV] & 67.0  & 29.8  & 31.8  & 33.8  & 35.7  \\
GMO $\;\; \:$ [MeV]               & 4.4   & 0.2   & 0.3   & 0.4   & 0.6   \\
\hline
\end{tabular}
\end{center}
\end{table}
\vskip 0.7cm

Table  1

\vskip 1.5cm

\begin{table}[bht]  \label{tab2}
\begin{center}
\begin{tabular}{l|c|c|c}
  & $\sigma_{\pi N} (2 m_{\pi}^2) - \sigma_{\pi N} (0) $   
  & $\sigma_{KN}^{(1)} (2 m_K ^2) - \sigma_{KN}^{(1)} (0) $ 
  & $\sigma_{KN}^{(2)} (2 m_K ^2) - \sigma_{KN}^{(2)} (0) $  \\
\hline
dim.                     & 7.4 & 318.9 $+\; i$ 334.9 & 9.4 $+\; i$ 334.9  \\
\hline
$\Lambda=300 \; (R_1)$   & 3.2 & 2.3 $-\; i$ 1.1 & -73.1 $+\; i$ 3.4  \\
$\Lambda=400 \; (R_1)$   & 3.8 & -22.8 $-\; i$ 7.4 & -107.5 $+\; i$ 22.2  \\
$\Lambda=500 \; (R_1)$   & 4.3 & 7.6 $-\; i$ 94.0 & -199.3 $+\; i$ 281.9  \\
$\Lambda=600 \; (R_1)$   & 4.6 & -220.1 $+\; i$ 373.4 & -1564 $+\; i$ 373.4\\
\hline
$\Lambda=300 \; (R_2)$   & 3.6 & -35.2 $+\; i$ 65.7 & 35.7 $-\; i$ 1.9  \\
$\Lambda=400 \; (R_2)$   & 4.1 & -84.8 $-\; i$ 127.6 & -45.5 $-\; i$ 266.7 \\
$\Lambda=500 \; (R_2)$   & 4.5 & -347.9 $+\; i$ 92.6 & -220.9 $-\; i$ 243.3\\
$\Lambda=600 \; (R_2)$   & 4.9 & -123.6 $+\; i$ 373.4 &-669.8 $+\; i$ 373.4\\
\hline
\end{tabular}
\end{center}
\end{table}
\vskip 0.7cm

Table  2

\newpage

\begin{table}[bht]  \label{tab3}
\begin{center}
\begin{tabular}{l|c|c|c|c|c}
  & dim. & $\Lambda=300$&$\Lambda=400$&$\Lambda=500$&$\Lambda=600$\\
\hline
$b_D \;\;$ [GeV$^{-1}$]             & 0.510 & 0.061 & 0.055 & 0.046 & 0.035 \\
$b_F \;\; $ [GeV$^{-1}$]            &-1.022 &-0.191 &-0.173 &-0.149 &-0.120 \\
$b_0 \;\; \:$ [GeV$^{-1}$]          &-1.591 &-0.424 &-0.351 &-0.264 &-0.169 \\
$\mnod_r \; \; $[GeV]           & 1.328 & 0.730 & 0.761 & 0.787 & 0.806 \\
$\sigma_{KN}^{(1)} (0)\; $ [MeV]  & -46.0 & 411.0 & 390.2 & 369.1 & 348.7 \\
$\sigma_{KN}^{(2)} (0)\; $ [MeV]  & -100.7 & 275.2 & 257.3 & 239.3 & 222.2 \\
SME $\; \; \;\:$ [MeV]            &-671.3 & 168.3 & 129.7 & 90.7 & 53.1 \\
$y$                               &-1.206 & 0.302 & 0.233 & 0.163 & 0.095 \\
$\hat{\sigma} \qq \q $ [MeV] & 99.3  & 29.8  & 31.4  & 34.5  & 40.7  \\
GMO $\;\; \:$ [MeV]               & 11.0   & 0.3   & 0.6   & 0.9   & 1.2   \\
\hline
\end{tabular}
\end{center}
\end{table}
\vskip 0.7cm

Table  3

\vskip 1.5cm

\begin{table}[bht]  \label{tab4}
\begin{center}
\begin{tabular}{l|c|c|c}
  & $\sigma_{\pi N} (2 m_{\pi}^2) - \sigma_{\pi N} (0) $   
  & $\sigma_{KN}^{(1)} (2 m_K ^2) - \sigma_{KN}^{(1)} (0) $ 
  & $\sigma_{KN}^{(2)} (2 m_K ^2) - \sigma_{KN}^{(2)} (0) $  \\
\hline
$\Lambda=400 \; (R_2)$    & 5.7 & -292.3 $+\; i$ 73.4 & -228.1 $-\; i$ 65.8 \\
$\Lambda=500 \; (R_2)$    & 6.5 & -534.8 $+\; i$ 425.7& -401.6 $+\; i$ 89.8 \\
\hline
\end{tabular}
\end{center}
\end{table}
\vskip 0.7cm

Table  4

\end{center}

\end{document}